\documentclass[aps,prl,twocolumn,amsmath,amssymb]{revtex4-1}

\usepackage{graphicx}
\usepackage{color}

\begin{document}


\title{Dimensionality constraints of light-induced rotation}


\author{L\'aszl\'o Oroszi$^1$, Andr\'as B\'uz\'as$^1$, P\'eter Galajda$^1$, L\'or\'and Kelemen$^1$, Anna Mathesz$^1$, Tam\'as Vicsek$^2$, Gaszton Vizsnyiczai$^1$ and P\'al Ormos$^1$}
\affiliation{
$^1$ Institute of Biophysics, Biological Research Centre, Hungarian Academy of Sciences, H-6726 Szeged, Hungary.\\
$^2$Department of Biological Physics, E\"otv\"os University, 
Statistical and Biological Physics Research Group of the Hungarian Academy of Sciences,\\
 H-1117 Budapest, Hungary.}


\date{\today}

\begin{abstract}
We have studied the conditions of rotation induced by collimated light carrying no angular momentum. Objects of different shapes and optical properties were examined in the nontrivial case where the rotation axis is perpendicular to the direction of light propagation. This geometry offers important advantages for application as it fundamentally broadens the possible practical arrangements to be realised.  We found that collimated light cannot drive permanent rotation of 2D or prism-like 3D objects (i.e. fixed cross-sectional profile along the rotation axis) in the case of fully reflective or fully transparent materials. Based on both geometrical optics simulations and theoretical analysis, we derived a general condition for rotation induced by collimated light carrying no angular momentum valid for any arrangement: Permanent rotation is not possible if the scattering interaction is two-dimensional and lossless. In contrast, light induced rotation can be sustained if partial absorption is present or the object has specific true 3D geometry. We designed, simulated, fabricated, and experimentally tested a microscopic rotor capable of rotation around an axis perpendicular to the illuminating light. 
\end{abstract}


\maketitle


Light-induced rotation has been an intensively studied effect ever since the availability of optical micromanipulation, partly because of the possible technical applications in micromechanics, at the same time the phenomenon is attractive for pure intellectual reasons, too \cite{1}. Different rotation mechanisms have been introduced and discussed which can be classified into two main categories. In one, light itself carries the angular momentum needed for the rotation (e.g. circularly polarized light \cite{2,3,4} Laguerre Gaussian beam \cite{4,5}, etc.). In the other scheme, light with no angular momentum is scattered by an object in a helical manner \cite{6,7,8,9,10} consequently, angular momentum is transferred to the object resulting in rotation (a process analogous to the rotation of a windmill). Here we discuss a yet not considered aspect of this latter case.\\\indent 
In the above mentioned cases rotation usually took place around an axis parallel to the optical axis. There is a distinctly different case, where the axis of rotation is perpendicular to the direction of illumination. This arrangement may be important for practical purposes: rotors could be arranged to roll in arbitrary directions on a plane illuminated perpendicularly by a single shared light beam (e.g. particles moving in independent directions on a flat surface could be supplied with energy by a single light source  – similar to the non-rotating wedge shaped quasy-autonomous robots introduced recently \cite{11}). Intuitively, one might think that this type of rotation should be possible simply by using properly designed asymmetric rotors. Indeed, water mill type light-driven rotors have been introduced earlier \cite{12,13,14}, however, in these cases rotation was achieved by partial illumination of the wheel (in analogy with a water mill partially submerged in water). In contrast, here we discuss fully illuminated rotors. According to our results, rotor geometries capable of sustained rotation are not trivial to construct. We conducted a systematic modelling study and theoretical analysis to determine the conditions for rotation and then experimentally verified our findings.\\\indent
In the modelling we studied objects of homogeneous material, either transparent or reflective (primarily non-absorbing, but absorption was also considered in specific cases). The illumination is a collimated unpolarised or linearly polarised light beam of homogeneous intensity distribution, fully covering the objects. In a geometrical optics approach, we calculated the forces acting on the rotors resulting from the linear momentum transfer between light and matter. Continuous illumination is modelled by the introduction of a high density array of light rays. Each ray is associated initially with unit light intensity. During the simulations these intensities are modified according to the Fresnel equations. When a light ray hits the interface of two regions having different refractive indices, then the reflected and refracted rays are created dynamically and new coordinates and intensities are calculated based on the law of reflection, Snell-Descartes law and the Fresnel equations in a recursive manner. The polarization state of the light is a freely adjustable parameter. The change in the angular momentum of light (with respect to some point of origin), thus the torque upon the object is a result of elementary linear momentum transfer processes integrated over all ray-object intersection points. We restrict ourselves to the following case (if not otherwise stated): (a) rotations are investigated around a fixed axis perpendicular to the illumination, (b) the illumination is modelled by a large number of homogeneously distributed, parallel, linearly polarized light rays, (c) mechanical interaction is based exclusively on transfer of linear momentum of light, (d) the motion is over-damped: the angular velocity is proportional to the torque exerted by light (e.g. microscopic particles in fluid).\\\indent
Two basic types of rotor shapes were investigated. First, we analysed prism like shapes, i.e. 3D objects that are created by the extrusion of a 2D footprint, and subsequently, we considered true 3D objects having a varying structure along the rotation axis. \\\indent
\begin{figure}[ht]
\includegraphics[width=0.45\textwidth]{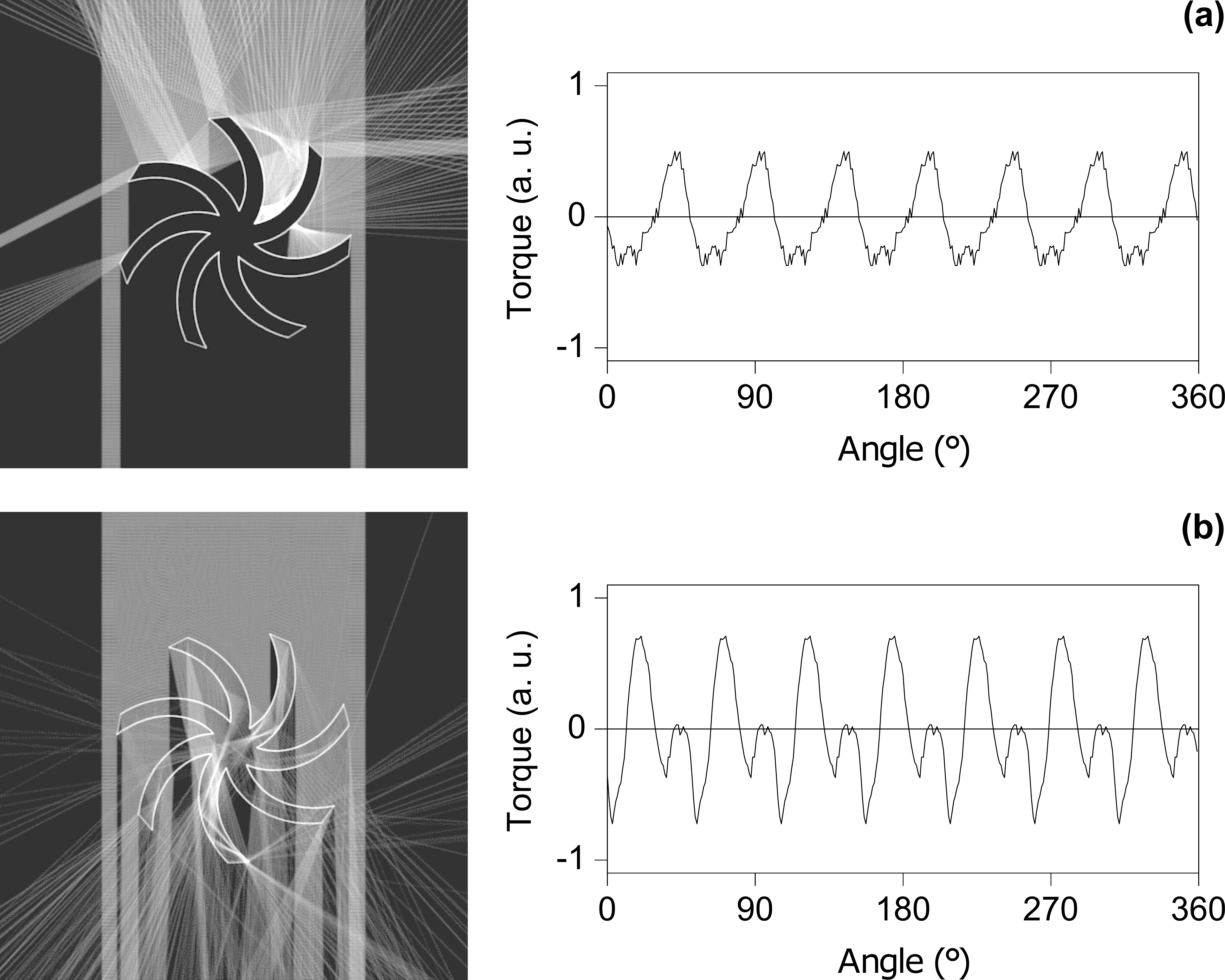}
\caption{Simulation of prism shape rotors in homogeneous collimated illumination perpendicular to the rotation axis.A characteristic rotor shape with reflective (a) and transparent (b) material and the respective torque exerted by light as a function of the rotation angle. In the images the reflected and refracted light rays are also presented.  Illumination comes from the top, the refractive index is 1.2.}
\label{fig1}
\end{figure}
In the case of prism-like rotors the problem is essentially two-dimensional: light scattering occurs in the plane perpendicular to the rotation axis. We investigated different shapes that one would intuitively regard as candidates for rotation. Fig. \ref{fig1} shows a rotor with characteristic shape made of reflective and transparent material interacting with collimated light. We have tried a large number of different shapes and we found that in the lossless case (no absorption) regardless of the shape of the object, either refractive or reflective, sustained rotation would not be driven by momentum transfer: Although in certain orientations light exerted torque upon the object, the integral of the torque for a full rotation cycle was always exactly zero, and in addition the object eventually ended up in an angular trap. \\\indent
The statement that prism shaped objects can not be rotated by collimated light can be proven with the following quantitative theoretical argumentation. The integral of torque for a full revolution of the object is equal to the torque experienced by the object at rest illuminated by light coming from all directions in the plane (i.e. isotropic illumination). This equivalent case is easier to analyse because the angular dependence has been removed. First, let us consider the case of the reflecting body. Any light beam with a given radiance (\(L=\frac{d^2P}{dA_{proj}\cdot d\Omega }\): power/projected area/steric angle) hitting the object will eventually leave it after a certain number of reflections, propagating away from the object in the plane of illumination. The radiance is known to be invariant for transfer in a lossless system: While during interaction with an optical surface parameters of the particular light beam may change, the radiance is conserved, and the outgoing beam will leave the object with unchanged radiance, independently of the distortions during particular refractions or reflections on surfaces. (We note that in our ray optics simulations this invariance is automatically fulfilled.)  Since the illumination is isotropic in the plane, the beam leaving the object will have its counterpart coming towards the object with exactly the same radiance (this is characteristic to the isotropic radiation field). It follows that the radiation field after the reflections will remain unchanged, consequently no net momentum change occurs, and therefore light will not have a mechanical effect upon the object. Thus we have also proven that for illumination from a single direction the integral of the torque for a full rotation is zero. \\\indent
The case of transparent objects can be understood using a similar argumentation. Let us consider an incident light beam of infinitesimally small cross section and radiance $\mathit{L}$. When hitting the surface of the object the beam is split into reflected and refracted beams typically several times before leaving the object, resulting in  multiple light paths. Traveling along a particular one of these paths the radiance of the incident light beam decreases after each transmission or reflection as determined by the Fresnel equations, giving an output radiance of $c_i\cdot\mathit{L}$. The original radiance is distributed among all possible light paths: $\sum{c_i}$=1.  Since the illumination is isotropic, light will also enter each path from the opposite direction with radiance $\mathit{L}$. Due to the direction invariance of the Fresnel coefficients the total attenuation of radiance will be the same in the two directions  (i.e. counterpropagating light will leave the path with output radiance $c_i\cdot\mathit{L}$ as well). Now, if we take into account all light paths into which a single incident light beam is divided, we can see that due to the direction invariance of the attenuation, the radiance of counterpropagating light will add up exactly to $\mathit{L}$, the original input radiance. This results again in an unchanged radiation field around the object, and thus no transferred torque.\\\indent
We have shown that in the two-dimensional case collimated light cannot induce sustained rotation of non-absorbing objects (either reflective or transparent). Interestingly, on the other hand, if losses are present, rotation becomes possible for certain rotor shapes. We present simulations of this case on rotors of the shape shown in  Fig. \ref{fig1}. In order to find the characteristic features of the phenomenon, we performed the simulations for different rotor geometries including varying number of blades (both odd and even) and calculated the torque exerted by light for the whole absorption range, as shown in Fig. \ref{fig2}. Assuming negligible inertia (e.g. microscopic objects in fluid) sustained rotation occurs only if the torque acting on the rotor does not change sign during a complete revolution (otherwise the object will be trapped in an angular trap). The condition of sustained rotation is satisfied only in certain absorption ranges specific to the actual rotor geometry. Note e.g. that in Fig. \ref{fig2}. in the case of the 5 blade rotor there is no absorptance region where the torque does not change sign, consequently, in this particular case continuous rotation does not take place at all. In other words, sustained rotation is possible only for specific combinations of absorption and rotor shape. Unfortunately, it is not possible to test this phenomenon experimentally, since in the case of significant absorption the sample would heat up to a level preventing the experiment.\\\indent 
\begin{figure}[ht]
\includegraphics[width=0.45\textwidth]{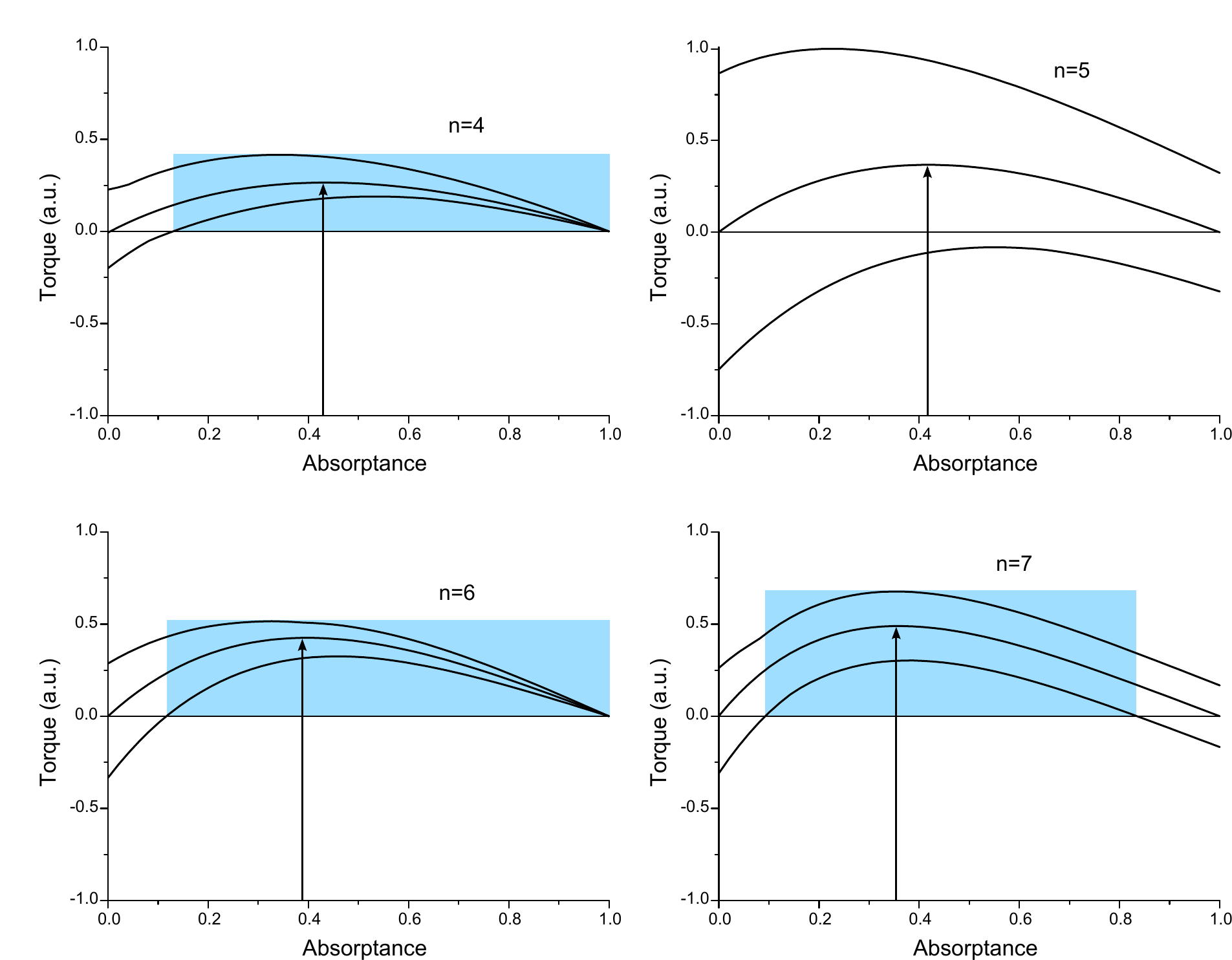}
\caption{Torque exerted by light on rotors of the shape shown in  Fig. \ref{fig1}.  with partially reflecting and partially absorbing surfaces. n is the order of the rotational symmetry of the object, i.e. the number of blades of the rotor. Average, minimum and maximum of the torque (evaluated over one revolution) is plotted as a function of absorptance. Steady rotation is possible only in certain regions of the absorptance where the torque acting on the rotor does not change sign during a complete revolution (marked by blue), specific for the rotor geometry. An arrow is positioned at the absorptance value where the average torque has its maximum.}
\label{fig2}
\end{figure}
\begin{figure}[ht]
\includegraphics[width=0.4\textwidth]{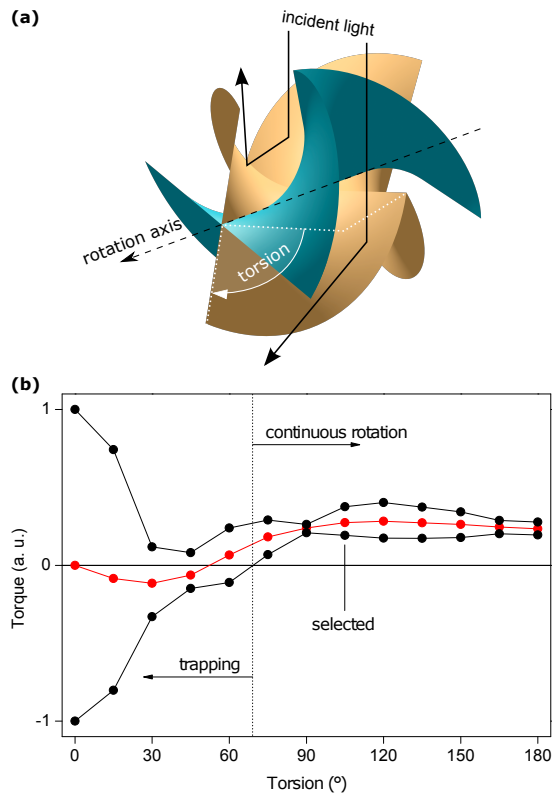}
\caption{Three-dimensional reflective rotor capable of sustained rotation around an axis perpendicular to a collimated light beam.  (a) Geometry of the rotor and illustration of the working principle. Torque results from the asymmetric momentum change of light on opposite sides of the rotation axis. The rotor has a simple geometry with one parameter, the torsion angle, and provides a smooth profile to enable rolling on a planar surface. (b) The average and the extreme values of the torque exerted by light as a function of the torsion angle. Sustained rotation is possible when the torque does not change sign during rotation (assuming negligible inertia, i.e. low Reynolds number regime). "Selected" indicates the torsion angle chosen for the realization of the rotor.}
\label{fig3}
\end{figure}
\begin{figure}[ht]
\includegraphics[width=0.45\textwidth]{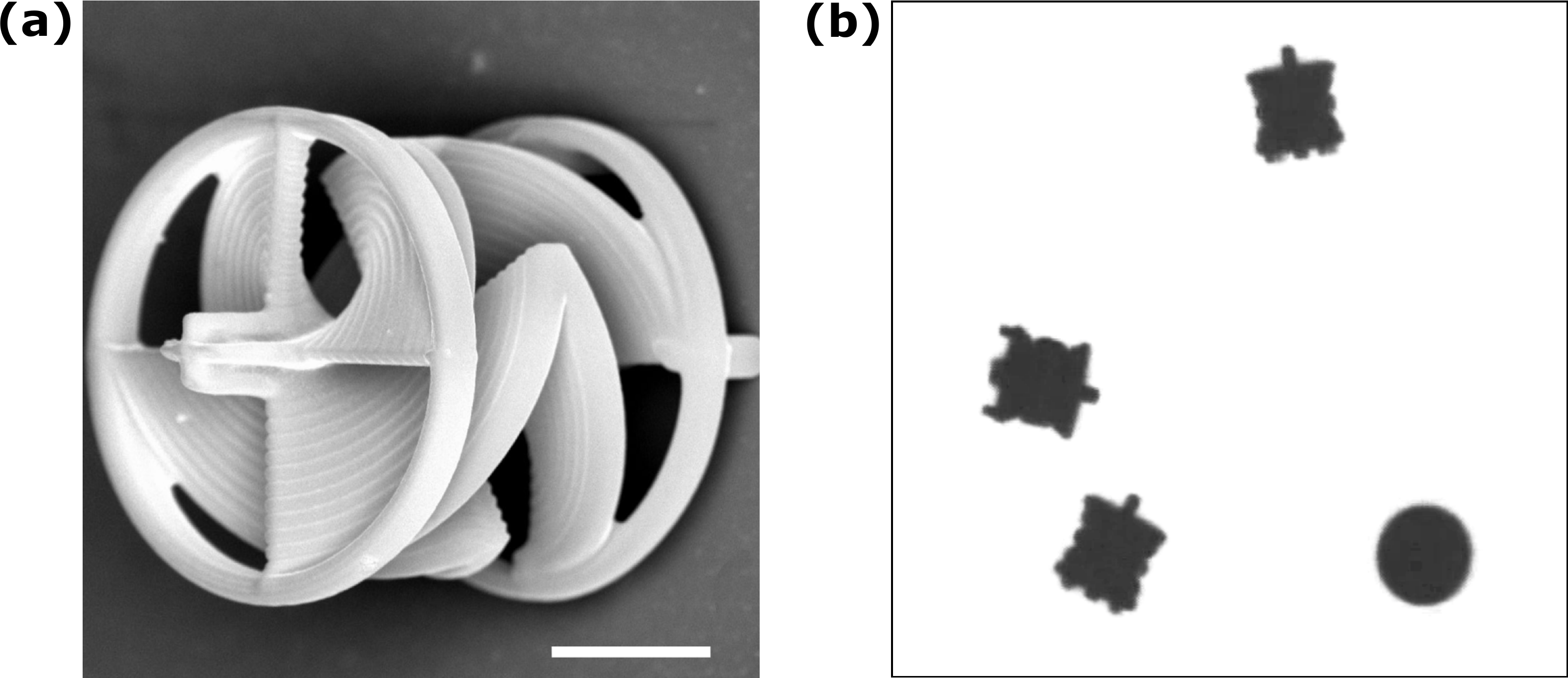}
\caption{The constructed rotor. (a) Electron microscopic image of the photopolymerized structure, the scale bar corresponds to 10 $\mu$m. (b) Optical image of the microrotors lying on the microscope slide. }
\label{fig4}
\end{figure}
The observation that permanent rotation is feasible if light is extracted (due to losses) from the two-dimensional plane where scattering occurs, suggests an additional possibility for rotation: deflection of light into the third dimension. To demonstrate this latter concept, we designed a true three-dimensional object where also the scattering process is three-dimensional. We chose a structure with four-fold symmetry, generated by the torsional deformation of a cross-based prism, twisted symmetrically in opposite directions along the two halves of the rotation axis (see Fig.\ref{fig3}.a.). The rotor fits into a cylinder, having identical height and diameter. With these constraints a single torsion angle parameter fully determines the geometry and consequently the torque exerted by light reflected form the surfaces. Fig.\ref{fig3}.b shows the results of the simulations, the capability of permanent rotation (i.e. minimum, maximum and average torque during a complete revolution) as a function of the torsion angle. The rotor can exhibit sustained rotation in a fairly wide range of torsion, thus making it more robust to potential fabrication errors. To experimentally confirm the results of the simulations we built a rotor with the essential features of the simulated object.\\\indent 
The rotors were produced by the two-photon photopolymerization technique \cite{8,15,16} from SU-8 photoresist (Michrochem, USA). After photopolymerization the surface of the rotors was made reflective at the wavelength of 1070 nm by covering them with a 200 nm thick Au metal mirror laye, on top of a 5 nm Cr layer for stability, both deposited on the surface by sputtering (Emitech K975X, UK). Fig. \ref{fig4} shows the fabricated microstructure.
\\\indent
In the experiments we realized rotation as rolling the rotor on a horizontal surface (microscope slide). The differences between the concept in Fig. \ref{fig3}. and the realization in Fig. \ref{fig4}. are due to practical necessities: Two rings were added at the edge of the structure to improve mechanical rigidity as well as to facilitate smooth rolling. Extruding tips were added to the rotation axes to ensure proper orientation of the rotors after sedimentation. 
\\\indent
The experiments were carried out in an inverted microscope (Zeiss Axio Observer A1). The sample compartment was formed by a microscope slide and cover slip separated by a 200$\mu$m spacer, and filled with water containing the rotors. After sedimentation, the rotors were rolling on the horizontal glass surface when illuminated vertically by an infrared fiber laser (IPG YLS-1070, I=10 W, $\lambda$=1070 nm). In addition to the illumination from the top we also applied illumination from the bottom with slightly less intensity: this scheme prevented the light pressing the rotor against the surface that could impede easy rolling of the rotors. To achieve this the linearly polarised light from the laser was passed through a $\lambda$/2 plate and then split into two beams by a polarising cube. One beam was directed to the sample compartment from below, one from above. The intensities of the two beams could be finely tuned by rotating the $\lambda$/2 plate. Upon illumination the rotors were rolling on the horizontal glass surface in directions defined by their actual orientations. Rotation took place as predicted by the simulations, the characterization of motion based on video analysis is shown in Fig.\ref{fig5}. (a typical video recording is included in the supplement at www.brc.hu/pub/lightdrivenrotor.avi). The rolling velocity was proportional to the illumination intensity, 100 W/cm$^2$ intensity corresponds to a velocity of 1.56 $\mu$m/s. Regular rolling of the rotors on the surface is demonstrated by the distribution of relative angles between travelling direction and rotor symmetry plane (Fig. \ref{fig4} b, also see video supplement).\\\indent
\begin{figure}[ht]
\includegraphics[width=0.45\textwidth]{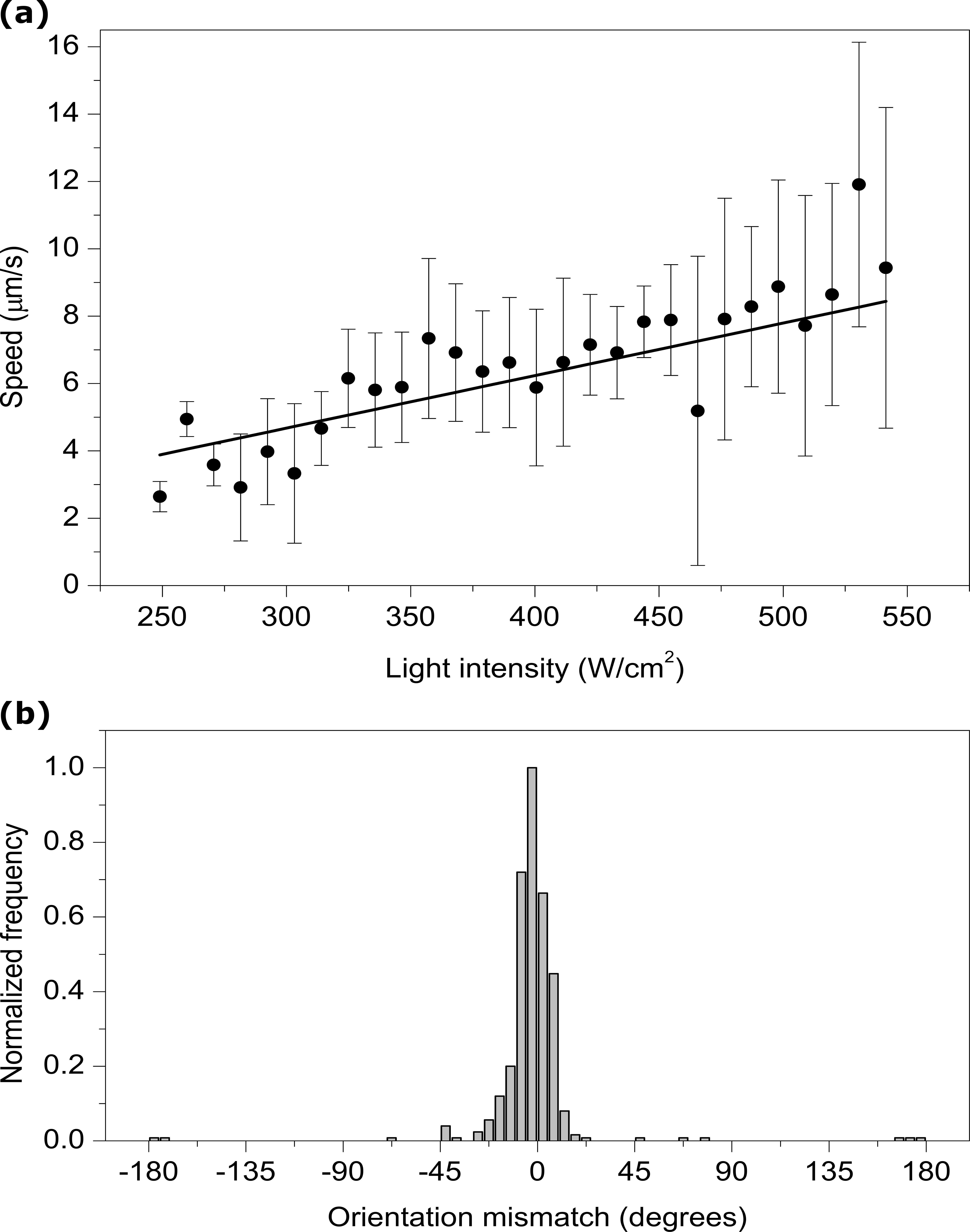}
\caption{Characteristics of microrotor motion. (a) Rotor speed as a function of the illumination intensity. The velocity values and the corresponding errors are determined by image analysis based on video recordings. The solid line represents a linear fit to the data points resulting in a slope of 0.0156 ($\mu$m/s)/(W/cm$^2$). (b) Histogram of the relative angle between the direction of rotor velocity and the symmetry plane of the rotor (based on the data set used in a).}
\label{fig5}
\end{figure}
We experimentally demonstrated that true 3D objects can be rotated by light in the discussed geometry. The crucial condition is that light is extracted from the scattering plane normal to the rotation axis by deflection of light into the third dimension (where the rotation axis points). To stress the peculiar nature of this characteristics, consider the cup anemometer that is rotated by wind. If it is reflective, it will also be rotated by light. If we take a two-dimensional version (projection) of it, simulations show that it will not rotate, in accordance with our findings concerning the dimensionality of the light scattering process. \\\indent
Finally we note that in all of the other arrangements for rotation discussed in previous works \cite{1,2,3,4,5,6,7,8,9,10}, where the rotation axis is parallel to (or has a component in) the direction of light propagation (propeller, windmill, etc.), the scattering process is three-dimensional. Taking all observations into account, we reach a general conclusion: Collimated light carrying no angular momentum cannot drive permanent rotation of any object (reflective or transparent) if the scattering process is two-dimensional and lossless.\\\indent
\begin{acknowledgments}
This work was supported by the grants OTKA NN102624,  NN114692, COST MP1205 and TAMOP-4.2.2.A-11/1/KONV-2012-0060 (to LO, AB, LK, AM, GV, PO) and EU ERC COLLMOT-227878 (to TV).
\end{acknowledgments}

\bibliography{myapsrev}

\begin{thebibliography}{}

\bibitem{1}
 M. Padgett and R. Bowman, Nature Photonics  {\bf5}, 343-348 (2011)

\bibitem{2}
M.E.J. Friese, T.A. Nieminen, N.R. Heckenberg and H. Rubinsztein-Dunlop,  Nature {\bf394}, 348 (1998)

\bibitem{3}
A.I. Bishop, T.A. Nieminen, N.R. Heckenberg and H. Rubinsztein-Dunlop, Physical Review A {\bf68}, 033802 (2003)

\bibitem{4}
N.B. Simpson, K. Dholakia, L. Allen and M.J. Padgett, Optics Letters {\bf22}, 52-54, (1997) 

\bibitem{5}
L. Paterson, M.P. MacDonald, J. Arlt, W. Sibbett, P.E. Bryant and K. Dholakia, Science {\bf292}, 912-914 (2001)

\bibitem{6}
A. Yamamoto and I. Yamaguchi, Jpn. J. Appl. Phys. {\bf34}, 3104-3108 (1995)

\bibitem{7}
R.C. Gauthier, Appl. Phys. Lett. {\bf67}, 2269-2271 (1995)

\bibitem{8}
P. Galajda and P. Ormos, Appl. Phys. Lett. {\bf78}, 249-251 (2001)

\bibitem{9}
T. Asavei, V.L.Y. Loke, M. Barbieri, T.A. Nieminen, N.R. Heckenberg and H. Rubinsztein-Dunlop, New Journal of Physics {\bf11}, (2009) 093021

\bibitem{10}
R. Di Leonardo, A. B\'uz\'as, L. Kelemen, G. Vizsnyiczai, L. Oroszi and P. Ormos, Phys. Rev. Lett. {\bf109}, 034104 (2012)

\bibitem{11}
A. B\'uz\'as, L. Kelemen, A. Mathesz, L. Oroszi, G. Vizsnyiczai, T. Vicsek and P. Ormos, Appl. Phys. Lett. {\bf101}, 041111 (2012)

\bibitem{12}
E. Higurashi, R. Sawada and T. Ito, Appl. Phys. Lett. {\bf72}, 2951 (1995) 

\bibitem{13}
R.C. Gauthier, R.N. Tait, H. Mende and C. Pawlowicz, Applied Optics {\bf40}, 930-937 (2001)

\bibitem{14}
L. Kelemen, S. Valkai and P. Ormos, Applied Optics {\bf45}, 2777-2780 (2006)

\bibitem{15}
S. Mauro, O. Nakamura and S. Kawata, Optics Lett. {\bf22} 132-134 (1997)

\bibitem{16}
L. Kelemen, S. Valkai and P. Ormos, Optics Express  {\bf15}, 14488-14497 (2007)

\end{thebibliography}

\end{document}